# Controlled low-temperature molecular manipulation of sexiphenyl molecules on Ag(111) using scanning tunneling microscopy

**Saw-Wai Hla**[*]
*Physics & Astronomy Department, Ohio University, Nanoscale and Quantum Phenomena Institute, Athens, OH 45701.*

**Kai-Felix Braun**
*Physics & Astronomy Department, Ohio University, Nanoscale and Quantum Phenomena Institute, Athens, OH 45701.*

**Bernhard Wassermann**
*Paul-Drude Institut für Festkörperelektronik, Hausvogteiplatz 5-7, D-10117 Berlin, Germany.*

**Karl-Heinz Rieder**
*Institut fuer Experimentalphysik, Freie Universitaet Berlin, D-14195 Berlin, Germany.*

A novel scanning tunneling microscope manipulation scheme for a controlled molecular transport of weakly adsorbed molecules is demonstrated. Single sexiphenyl molecules adsorbed on a Ag(111) surface at 6 K are shot towards single silver-atoms by excitation with the tip. To achieve atomically straight shooting paths, an electron resonator consisting of linear standing wave fronts is constructed. The sexiphenyl manipulation signals reveal a π-ring flipping as the molecule moves from hcp to fcc site. Abinitio calculations show an incorporation of the Ag atom below the center of a π-ring.

[*]Coressponding author, Email: hla@ohio.edu,
web: www.phy.ohiou.edu/~hla

---

The advances in scanning tunneling microscope (STM) manipulation allow probing physical/chemical properties of single molecules or construction of atomic scale structures on surfaces [1-9]. STM manipulation requires a precise control over the tip-molecule-surface junction. A weakly adsorbed molecule on a surface can be easily displaced with the STM-tip but its movement is extremely difficult to control. Most surface chemical reactions involve weakly adsorbed molecular species; however, investigations on their detailed dynamics and reactivity are hindered by instrumentation limits. Here, we have chosen weakly adsorbed sexiphenyl on Ag(111) as a model system to develop an STM manipulation scheme. Sexiphenyl ($C_{36}H_{26}$) is composed of six π-rings connected to form a linear chain [10] and due to its potential applications in display electronic devices, sexiphenyl has been studied intensely in the past years [10-15, 21].

The experiments were performed by using an ultra-high-vacuum low-temperature STM at 6 K. The Ag(111) sample was cleaned by sputter-anneal cycles. A minute amount of sexiphenyl was deposited onto the sample at ~70 K by thermal evaporation and the sample was cooled down to 6 K for the STM experiment. STM images show that molecules mostly position with their long molecular-axis parallel to the surface close-packed atom rows. Because of the weak molecule-substrate binding, sexiphenyl molecules are easily displaced during imaging and they are often dragged along with the STM-tip [24]. Atomically controlled manipulation of sexiphenyl [6] is extremely difficult; the molecules orient randomly and occasionally slip away from the STM-tip.

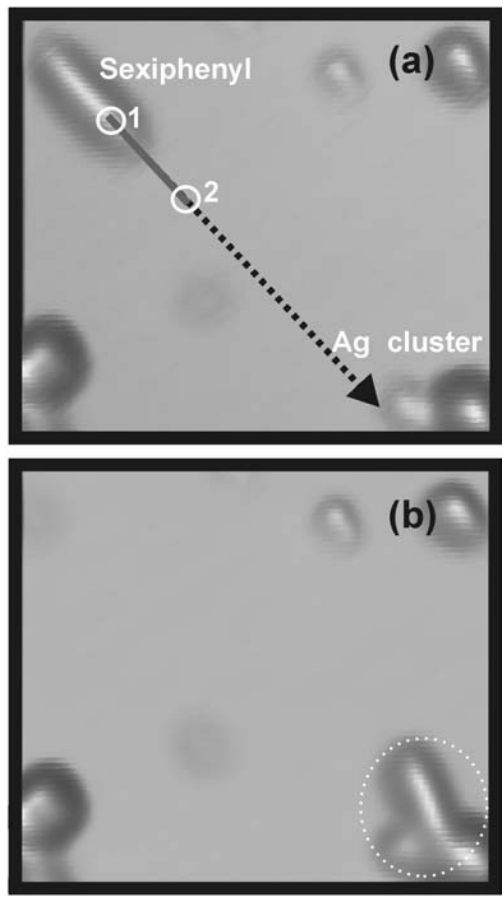

**FIG. 1.** *Molecule propagation. (a) A sexiphenyl molecule at the upper left corner is dragged over 3.3nm along the solid line from location '1' to '2'. (b) The molecule continues to travel further after withdrawing the tip and hits a silver cluster located at the lower right corner 10 nm away from the initial position. The molecule's final location deviates by ~2nm from the straight-line path indicated by the dashed arrow in (a). [Image size: 11 nm x 9.5 nm, shooting parameters: $R_t = 1.5 \times 10^5\ \Omega$, $V_t = 30$ mV, molecule dragging parameters: $V_t = 20$ mV, $R_t = 4 \times 10^4\ \Omega$, [22]].*

To 'shoot' the sexiphenyl, the STM-tip is approached towards the molecule and then dragged the molecule for a few nm [Fig. 1(a)]. When the tip releases the molecule by retracting back to the imaging height, the molecule propagates further across the surface [Fig. 1(b)]. Repeating the procedure shows up to 30 nm propagations, which is the average distance to encounter a defect/step-edge on our sample. Such „shooting" propagation occurs only when sexiphenyl is manipulated along its long-molecular axis in the close-packed surface directions. Attempts to precisely shoot the molecules are not successful;

their propagation paths mostly deviate from the atomically straight-lines [Fig.1(b)]. Sexiphenyl is sensitive to the local surface electronic structural environment, such as random electron standing waves produced by defects, causing the molecule to deflect from its original propagation path. The electron standing waves are known to influence the adsorbate diffusion via long-range attractive/repulsive interactions [16-19].

To achieve atomically straight-line propagation paths, an electron resonator exhibiting linear standing-wave fronts is constructed by using silver atoms extracted from the native substrate. This process represents a *novel atomic-scale construction concept*: The STM-tip is initially dipped into the substrate producing a crater on the surface, around which nanometer sized clusters are scattered [Fig. 2(a)]. Individual silver atoms are then pulled out from the clusters using the tip-atom attractive interaction [9]. These atoms are then relocated with the STM-tip to form two one-dimensional atom arrays aligned along a surface close-packed direction with six nearest-neighbor atom distance (1.734 nm) separation [Fig. 2(b)].

To shoot the sexiphenyl molecules, two individual silver atoms (targets) and two sexiphenyl molecules are positioned at opposite ends of the resonator in atomically straight-line paths [Fig. 3]. The molecules are then shot toward the targets by employing the procedure described above. Now, the sexiphenyl propagates in an atomically straight-line path and hits the target atom thereby a sexiphenyl-silver complex is formed [Fig. 3]. Both, shooting along the standing wave minima and the maxima yield the same results [24] indicating that the uniform electronic-structural environment parallel to the linear standing wave fronts is the key to achieve atomically straight molecule propagation. The stability of the complex is examined by lateral manipulation [9]: The entire unit can be moved back and forth without losing the atom underneath revealing a strong molecule-adatom interaction.

We have performed a DFT computation with LDA (DGauss 5.0 program package) [12] using the Gaussian basis set DZVP to study the complex structure in the absence of the substrate. The STM height profile of the complex [Fig. 3(c)] reveals that only half of the molecule is distorted to enclose the silver atom. Hence, only three π-rings have been used in computation. Among several molecule-atom geometries [24] the minimum energy structure reveals bending of three π-rings to enclose the atom thereby enhancing the molecule-Ag interaction [Fig. 3(d)], in well agreement with the measurement [Fig. 3(c)].

Fig. 4(a) shows the molecule-substrate registry determined from the atomically resolved STM images and from the silver atom positions of the resonator [20]. Sexiphenyl aligns the surface close-packed directions with the alternately twisted π-rings [15,20,21], as in the gas phase [21] and on the Al(111) surface [15], in agreement with the weak adsorption.

From the measured atomic registry [Fig. 4(a)], it is apparent that shifting the molecule for a half of the nearest-neighbor silver atom distance forward, i.e. hcp-fcc sites, will flip the π-rings from one tilting position to another [20].

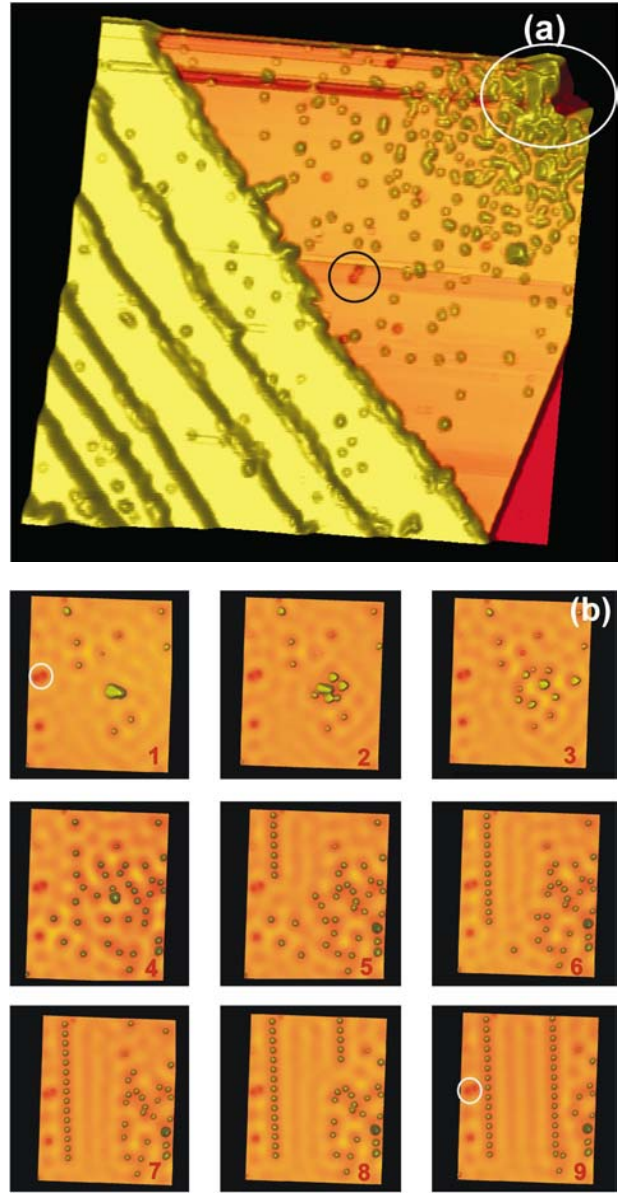

*FIG. 2 (color). Atomic scale Construction: A nanometer size hole (circled at top-right corner) with a large number of Ag clusters surrounding it has been formed by controlled dipping of the STM-tip into the Ag(111) substrate (a). The circle at the center indicates two unknown adsorbates used here as a landmark. Several sexiphenyl molecules, which were deposited earlier, are sticking at the step-edges. (b) Snap-shot STM images from an STM movie [24] show the atom detachment from a cluster and the atom-by-atom construction of the electron resonator. A white circle indicates the landmark in the first and the last image. Notice that the random background standing wave pattern in '1' becomes a parallel uniform pattern in '9'. The distance between two standing wave minima is 3.9 nm, half of the Fermi wavelength of the Ag(111) surface state [6]. This quantum structure is 25 nm long, 12 nm wide and consists of 30 silver atoms. [Imaging parameters: $V_t = 39$ mV, $I_t = 1.1$ nA].*

track all the way inside the electron resonator in a constant-current mode, where the tip-height is maintained by the feedback loop. The recorded tip-height signal [Fig. 4(b)] shows two unequal-height contours periodically repeating at 0.145 nm, half of the nearest-neighbor silver atom distance. This reveals the π-ring flipping at every hcp-fcc site as follows: When the STM-tip encounters the down site of the π-ring [Fig. 4(c)], it produces a lower height signal while the up-site of the π-ring causes the higher height signal [Fig. 4(d)].

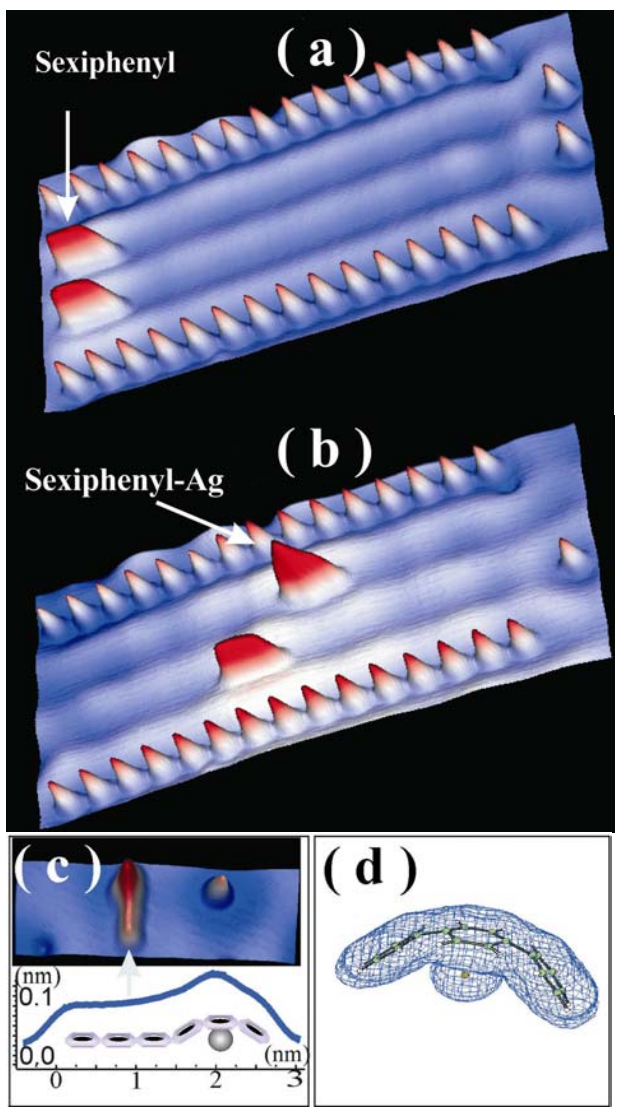

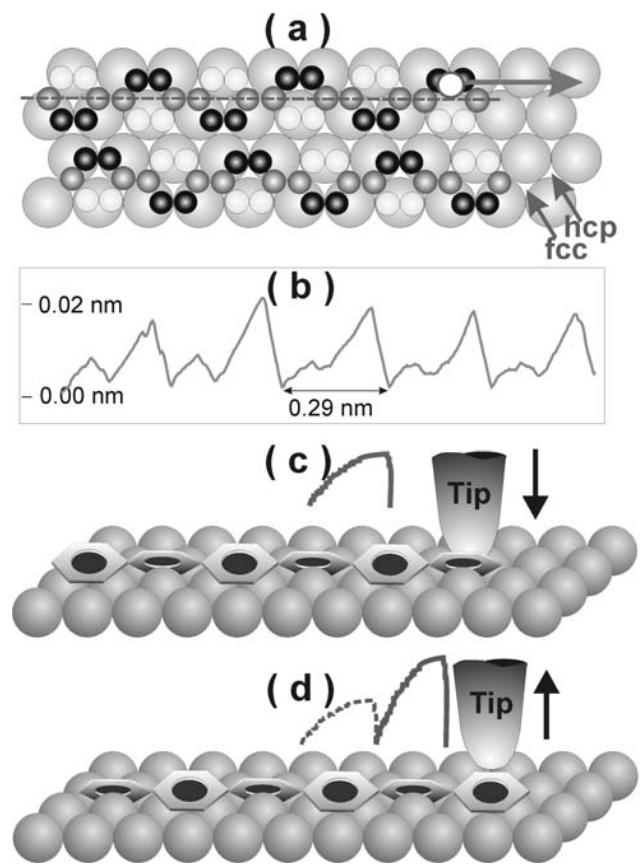

**FIG. 3 (color).** *Controlled molecular shooting: Two sexiphenyl molecules (left end) and two silver atom targets (right end) are positioned along the standing wave track (a). A silver-sexiphenyl complex is formed by shooting the upper sexiphenyl with the STM-tip. (b) Both the complex (upper) and the bare sexiphenyl (lower) are laterally moved with the STM-tip to the middle of the electron resonator to ease for a visual comparison. (c) Upon hitting the target, the sexiphenyl locates above the target atom by bending π-rings. The inset shows a measured STM-tip height profile along the long molecular axis of the complex (indicated with the light-color arrow). The inset drawing illustrates the measured silver atom position relative to the π-rings. (d) The computed silver-tarphenyl complex structure confirms bending of π-rings from the inter-ring joints to enclose the silver atom. [Imaging parameters: $V_t = 30$ mV, $I_t = 1.1$ nA, $16 \times 26$ nm$^2$].*

To understand the *molecule propagation mechanism*, we directly measure the internal-conformation changes of sexiphenyl by laterally moving it with the STM tip [9]. To observe the π-ring movement, the STM-tip is positioned above a π-ring-edge and drags the molecule along the standing-wave

**FIG. 4.** *Molecular conformation changes of sexiphenyl: (a) Sexiphenyl preferentially align along the surface close-packed directions. The dotted line indicates the long molecule axis. The π-ring edge is tilted up when the two carbon atoms sit on top of a single surface atom (light balls) and tilted down when they sit on two surface atoms (dark balls) [20]. Moving the molecule (upper drawing) for a half–atomic distance forward (hcp-fcc sites) will switch the π-ring position relative to the surface atoms (lower drawing). The up-site π-ring edge will now become down-site and vice-versa resulting in the π-ring flipping [20]. The tip is positioned (white circle) 0.27 nm above the π-ring edge and moved along the arrow pointed direction. (b) Periodic low-high peak tip-height signal repeating at every half of the silver atom distance is observed during LM. (c) A low-height manipulation signal is recorded when the tip is in the low-site of the π-rings; the higher height signal is obtained when the π-ring edge is lifted up (d) (as indicated by the arrows) [Manipulation parameters: $V_t = 49$ mV, $R_t = 600$ kΩ, [22]].*

During the shooting process, the surface phonon contributions to the sexiphenyl propagation are assumed to be negligible because of the low temperature and relatively large mass of the molecule. Due to the weak molecule-substrate binding, the molecule can be assumed as a relatively isolated system. Klafter and co-workers [23] have proposed a model to transport a one-dimensional atomic-chain across a surface, which involves translational motion and changing atomic bond-lengths: The potential energy gained in the bond-length changes is transferred to the kinetic energy that in turn changes the bond-length with the help of the substrate potential leading to transport the entire atom-chain across the surface. In our case, the π-ring flipping is associated with the surface atomic-geometry. As described above, the sexiphenyl propagation will cause the π-ring flipping, which in turn will cause translational motion of the molecule with the help of the periodic surface-potential landscape. Here, the sexiphenyl propagates by flipping the π-rings to and fro - like a worm traveling on ground [24]- while the energy is transferred back and forth between the potential energy (π-ring flipping) and the kinetic energy (the translational motion). The one-dimensional geometry of the molecule and the relatively matching of periodic surface-potential landscape with the molecule length along the surface close-packed directions allow this mechanism to occur. Variation of both the dragging distance (from 0.5 to 4 nm) and the tip-speed (0.2 to 1 nm/sec) yield the same shooting results indicating that these are not the main cause. The mechanical energy supplied by the STM-tip is sufficient to move the molecule. If the tip releases the molecule where it is not at the saturated position, the molecule will need to move forward to achieve a full swing of the π-ring tilting. This could trigger the entire shooting process.

In summary, we have demonstrated atomically precise manipulations of weakly adsorbed sexiphenyl molecules using an artificial manipulation environment that is also constructed locally. Even though the entire experiment was conducted within an area of 40 x 40 nm$^2$, atomic-level information concerning the molecular conformation changes and the selective reactivity of sexiphenyl towards silver adatom/surface could be obtained. Our experiment not only reveals the extent of the surface electronic structure influence on the propagation of a physisorbed molecule but also we have applied this phenomenon further to develop a molecular shooting scheme. The potential applications of our achievement include induction of $SN_2$ type reactions and molecular transport across the surface.

This work is supported by the US Department of Energy, BES DE-FG02-02ER46012, by Deutsche Sfb 290 and by EU-EFRE grants. We gratefully acknowledge N. Koch, A. Volmer, H. Castillo and J.R. Manson for insightful discussions.

**References**
[1]. J.A. Heinrich, C.P. Lutz, J.A. Gupta, J.A. D.M. Eigler, Science **298**, 1381 (2002).
[2]. C. Joachim, J.K. Gimzewski, A. Aviram, Nature **408**, 541 (2000).
[3]. H.C. Manoharan, C.P. Lutz, D.M. Eigler, Nature **403**, 512 (2000).
[4]. F. Rosei et al. Science **296**, 328 (2002).
[5]. S.-W. Hla, L. Bartels, G. Meyer, K.-H. Rieder, Phys. Rev. Lett. **85**, 2777 (2000).
[6]. S.-W. Hla, K.-H. Rieder, Ann. Rev. Phys. Chem. **54**, 307 (2003).
[7]. G. Dujardin, A.J. Mayne, F. Rose, Phys. Rev. Lett. **89**, 036802-1 (2002).
[8]. Z.J. Donhauser et al. Science **292**, 2303 (2001).
[9]. S.-W. Hla, K.-F. Braun, K.-H. Rieder, Phys. Rev. B. **67**, 201402-1R (2003).
[10]. M. Knupfer, E. Zojer, G. Leising, J. Fink, Synth Met. **119**, 499 (2001).
[11]. Y.Z. Wang, et al. Synth. Met. **102**, 889 (1999).
[12]. N. Koch, et al. Surf. Sci. **454-456,** 1000 (2000).
[13]. A. Kahn, N. Koch, W. Gao, J. Pol. Sci. B **41**, 2529 (2003).
[14]. M.B. Johnston et al. Chem. Phys. Lett. **377**, 256 (2003).
[15]. J. Ivanco, B. Winter, F.P. Netzer, M.G. Ramsey, Adv. Mater. **15**, 1812 (2003).
[16]. J. Repp et al. Phys. Rev. Lett. **85**, 2981 (2000).
[17]. M.M. Kamna, S.J. Stranick, P.S. Weiss, Science **274**, 118 (1996).
[18]. S. Lukas, G. Witte, Ch. Wöll, Phys. Rev. Lett. **88**, 028301-1 (2001).
[19]. N. Knorr et al. Phys. Rev. B **65**, 115420-1 (2002).
[20]. K.F. Braun, S.-W. Hla, submitted for publication.
[21]. B. Champagne, D.H. Mosley, J.G. Fripiat, J.M. Andre, Phys. Rev. B **54**, 2381 (1996).
[22] Instead of tunneling current ($I_t$), the tunneling resistance ($R_t$) is used to indicate the tip-sample distance and the tip-sample interaction strength [6] in STM lateral manipulation experiments.
[23]. M. Porto, M. Urbakh, J. Klafter, Phys. Rev. Lett. **84**, 6058 (2000).
[24]. See EPAPS document for STM movies: Atomic-scale construction, sexiphenyl movement. A direct link to this document may be found in the online article's HTML reference section.